\documentclass[12pt,preprint]{aastex}

\slugcomment{Accepted for publication in ApJL}

\shorttitle{Star formation in the host of the Cloverleaf QSO}
\shortauthors{Lutz et al.}

\begin{document}

\title{PAH emission and star formation in the host of the z$\sim$2.56 
Cloverleaf QSO}


\author{D. Lutz\altaffilmark{1},
E. Sturm\altaffilmark{1}, L.J. Tacconi\altaffilmark{1}, E. Valiante\altaffilmark{1}, M. Schweitzer\altaffilmark{1}
H. Netzer\altaffilmark{2},
R. Maiolino\altaffilmark{3}, 
P. Andreani\altaffilmark{4}, 
O. Shemmer\altaffilmark{5}, 
S. Veilleux\altaffilmark{6}}
\altaffiltext{1}{Max-Planck-Institut f\"ur extraterrestrische Physik,
Postfach 1312, 85741 Garching, Germany \email{lutz@mpe.mpg.de, 
sturm@mpe.mpg.de, linda@mpe.mpg.de, 
valiante@mpe.mpg.de, schweitzer@mpe.mpg.de}} 
\altaffiltext{2}{School of Physics and Astronomy and the Wise Observatory,
      The Raymond and Beverly Sackler Faculty of Exact Sciences,
     Tel-Aviv University, Tel-Aviv 69978, Israel\email{netzer@wise1.tau.ac.il}}
\altaffiltext{3}{INAF, Osservatorio Astronomico di Roma, via di Frascati 13,
   00040 Monte Porzio Catone, Italy\email{maiolino@ao-roma.inaf.it}}
\altaffiltext{4}{ESO, Karl-Schwarzschildstra\ss{}e 2, 85748 Garching, Germany
  \email{pandrean@eso.org}}
\altaffiltext{5}{Department of Astronomy and Astrophysics, 525 Davey 
Laboratory, Pennsylvania State University, 
   University Park, PA 16802, USA\email{ohad@astro.psu.edu}}
\altaffiltext{6}{Department of Astronomy, University of Maryland, 
      College Park, MD 20742-2421, USA\email{veilleux@astro.umd.edu}}

\begin{abstract}
We report the first detection of the 6.2$\mu$m and 7.7$\mu$m 
infrared `PAH' emission features in the spectrum of a high redshift QSO, 
from the Spitzer-IRS spectrum of the Cloverleaf lensed QSO (H1413+117, 
z$\sim$2.56).
The ratio of PAH features and rest frame far-infrared emission is the same 
as in lower luminosity star forming ultraluminous infrared galaxies and in 
local PG QSOs, supporting a predominantly starburst nature of the Cloverleaf's 
huge far-infrared luminosity ($5.4\times 10^{12}$L$_\odot$, corrected for 
lensing). The Cloverleaf's period of dominant QSO 
activity (L$_{Bol}\sim 7\times 10^{13}$L$_\odot$) is coincident with 
an intense (star formation rate $\sim 1000$M$_\odot$yr$^{-1}$) and short (gas
exhaustion time $\sim 3\times 10^7$yr) star forming event.

\end{abstract}


\section{Introduction}
Redshifts $\sim$2.5 witness both the `quasar epoch' with
peak number density of luminous accreting black holes 
\citep[e.g.][]{schmidt95} 
and the peak in number of the most intense star forming
events as traced by the submillimeter galaxy population \citep{chapman05}, 
suggestive of a relation of the two 
phenomena. Detailed evolutionary connections between massive starbursts 
and QSOs have been discussed for many years 
\citep[e.g.][]{sanders88,norman88} and form an
integral part of some recent models of galaxy and merger evolution 
\citep[e.g.][]{granato04,springel05,hopkins06}. Phases of intense star 
formation coincident with the active phase of the quasars are a natural 
postulate of such models but have been exceedingly difficult 
to confirm and quantify due to the effects of the powerful
AGN outshining tracers of star formation at most wavelengths. 

Perhaps the strongest constraint on the potential significance of star 
formation in QSOs comes from the far-infrared part of their spectral energy 
distribution (SED). Indeed, this far-infrared emission has been interpreted as due
to star formation 
\citep[e.g.][]{roro95}, but alternative models successfully ascribe it
to AGN heated dust, by postulating a dust distribution in which 
relatively cold dust at large distance from the AGN has a significant covering
factor, for example in a warped disk configuration \citep{sanders89}. 
Additional diagnostics are needed to break this degeneracy.

CO surveys of local QSOs \citep[e.g.][]{evans01,scoville03,evans06} have 
produced a 
significant number of detections of molecular gas reservoirs that might
power star formation. Depending on the adopted `star formation efficiency'
SFE=L$_{FIR}$/L$_{CO}$ the detected gas masses may be sufficient or not
for ascribing the QSO far-infrared emission to star formation. Optical studies
have identified significant `post-starburst' stellar
populations in QSOs \citep{canalizo01,kauffmann03}. On the other hand, 
\citet{ho05} suggested low star formation in QSOs, perhaps actively 
inhibited by the AGN, on the basis of observations of the 
[OII] 3727\AA\ line. We have used
the much less extinction sensitive mid-infrared PAH emission features to
infer that in a sample of local (PG) QSOs, star formation is sufficient
to power the observed far-infrared emission \citep{schweitzer06}.

The observational situation remains complex for high redshift QSOs. 
Metallicity studies of the broad-line region
suggest significant enrichment by star formation 
\citep[e.g.][]{hamann99,shemmer04} but
may not be representative for the host as a whole.
Submm and mm studies of luminous radio quiet QSOs have
produced significant individual detections of dust emission of 
some QSOs, as well as statistical detection
of the entire population \citep[e.g.][]{omont03,priddey03,barvainis02}.
These suggest potential starburst luminosities up to and exceeding 
10$^{13}$L$_\odot$. CO studies have detected large gas reservoirs in many 
high-z QSOs \citep[see summaries in][]{solomon05,greve05}. Emission from
high density molecular gas tracers has been detected in some of the brightest
systems \citep{barvainis97,solomon03,carilli05,riechers06,garcia06,guelin07} 
and may well originate in dense high pressure star forming 
regions, but AGN effects on chemistry and molecular line excitation 
could also play a role \citep[e.g.][]{maloney96}.
Finally, the [CII] 157$\mu$m rest wavelength fine structure line 
was detected in the z=6.42 quasar 
SDSS J114816.64+525150.3 \citep{maiolino05} at a ratio to the rest frame 
far-infrared emission similar to the ratio in local 
ULIRGs, consistent with massive star formation.

We have initiated a program extending the use of mid-infrared PAH emission 
as star formation tracer to high redshift QSOs.
In this Letter, we use Spitzer mid-infrared spectra to detect and quantify 
star formation in one of the brightest and best studied z$\sim$2.5 QSOs, the 
lensed Cloverleaf \citep[H1413+117,][]{hazard84,magain88}. We adopt 
$\Omega_m =0.3$, $\Omega_\Lambda =0.7$ and $H_0=70$ km\,s$^{-1}$\,Mpc$^{-1}$.

\section{Observations and Results}
We obtained low resolution (R$\sim 60-120$) mid-infrared spectra of the 
Cloverleaf QSO using the Spitzer infrared spectrograph IRS \citep{houck04} in 
staring mode on July 24, 2006, at
J2000 target position RA 14h15m46.27s, DEC +11d29m43.40s. The IRS aperture
includes all lensed images. 30 cycles of 
120sec integration time per nod position were taken in the 
LL1 (19.5 to 38.0 $\mu$m) and 15 
cycles in the LL2 (14.0 to 21.3 $\mu$m) module, leading to effective on-source 
integration times of 2 and 1 hours, respectively. We use the pipeline 14.4.0 
processed basic calibrated data, own deglitching and coaddition procedures, 
and SMART \citep{higdon04} for extraction.
When combining the two orders
into the final spectrum, we scaled the LL2 spectrum by a factor 1.02 for 
best match in the overlapping region. 

Fig.~\ref{fig:sed} shows the IRS spectrum embedded into the infrared to radio 
SED of the Cloverleaf, and Fig.~\ref{fig:irs} 
the IRS spectrum proper, together with the location of key features in the
corresponding rest wavelength range. The rest frame mid-infrared emission is 
dominated by a strong continuum, approximately flat in $\nu$F$_\nu$, due to 
dust heated by the powerful active nucleus to temperatures well above
those reached in star forming regions. Superposed on this 
continuum are emission features, which we identify with the 6.2$\mu$m and 
7.7$\mu$m aromatic `PAH' emission features normally detected in 
star-forming galaxies
over a very wide range of properties. As expected for a Type 1 AGN, there
 are no indications for the ice (6$\mu$m) or silicate (9.6$\mu$m) absorptions
seen in heavily obscured galaxies. None of the well-known emission 
lines in this wavelength range is bright enough to be significantly 
detected in this low resolution spectrum, although we cannot
exclude a contribution of [Ne\,VI] 7.64$\mu$m to the 7.7$\mu$m feature.
Adopting standard mid-infrared low resolution diagnostics 
\citep{genzel98,laurent00}, the weak PAH features on top of a strong continuum
agree with the notion that the Cloverleaf is energetically dominated by
its AGN. The detection of PAH features with several mJy peak flux density in a 
z$\sim$2.6 galaxy, however, implies intense star formation, which we discuss
in conjunction with other properties of the Cloverleaf.

By fitting a Lorentzian superposed on a local polynomial continuum, we measure
a flux of 1.52$\times 10^{-21}$W\,cm$^{-2}$ for the 6.2$\mu$m feature, with a 
S/N of 6. The 7.7$\mu$m feature is more difficult to quantify. 
\citet{schweitzer06} have discussed PAHs as star formation indicators in
local (PG) QSOs, PAHs are also detected in the average QSO 
spectrum of \citet{hao07}. The AGN continuum of those 
QSOs shows superposed silicate 
emission features at $\gtrsim 9\mu$m \citep[see also][]{siebenmorgen05, hao05}.
If PAH emission is additionally present, the PAH features partly 
`fill in' the minimum
in the AGN emission before the onset of the silicate feature (see 
Fig.~2 of \citet{schweitzer06}), causing a seemingly flat overall spectrum.
In reality, there is simultaneous presence of AGN continuum, the 
6.2-8.6$\mu$m PAH complex, silicate emission, and more PAH emission at
longer wavelengths. From inspection of 
Fig.~\ref{fig:irs}, a similar co-presence of PAH and silicate emission is
observed for the Cloverleaf. We note that the presence of silicate emission
in the luminous Cloverleaf Type 1 QSO, other high z Type 1 QSOs 
(Maiolino et al., in prep) as well as in luminous Type 2 QSOs 
\citep{sturm06, teplitz06} has implications for the location of 
this cool silicate component \citep[$\sim$200K,][]{hao05} in unified AGN 
schemes. We leave further discussion of
the properties of silicates to a future paper. For the 7.7$\mu$m feature
of the Cloverleaf, we adopt a flux of 6.1$\times 10^{-21}$W\,cm$^{-2}$. This 
flux was determined by scaling a PAH template 
\citep[ISO spectrum of M82,][]{sturm00} to the measured Cloverleaf 
6.2$\mu$m feature flux, and then fitting three lorentzians to represent the 
6.2, 7.7, and 8.6 features plus a local polynomial continuum to this template.
Similar Lorentzian fits were also used for local comparison objects discussed 
below.
\citet{brandl06} have quantified the scatter of the 6.2 to 7.7$\mu$m flux ratio
in starbursts, with 0.07 in the log this scatter indicates the modest 
uncertainty 
induced by tying the longer wavelength features to the 6.2$\mu$m one.
The result of subtracting the scaled M82 template from the Cloverleaf 
spectrum is indicated in Fig.~\ref{fig:irs}, and shows a combination of
continuum and silicate emission very similar to local QSOs.
Directly measuring the 7.7$\mu$m flux by fitting a single Lorentzian plus 
local continuum to the Cloverleaf spectrum gives a $\sim$40\% lower feature
flux, which would be a systematic underestimate because of the complexity
of the underlying continuum/silicates discussed above.

\section{Intense star formation in the host of the Cloverleaf QSO}

The Cloverleaf SED (Fig.~\ref{fig:sed}) shows strong rest frame far-infrared
emission in addition to the AGN heated dust emitting in the rest frame
mid-infrared. \citet{weiss03} decomposed the SED into two modified blackbodies
of temperature 50 and 115\,K, the rest frame far-infrared (40-120$\mu$m) 
luminosity of $5.4\times 10^{12}$L$_\odot$ is dominated by the colder component
and could largely originate in star formation. Comparison of PAH and 
far-infrared emission can shed new light on this question. The bolometric
(rather than rest-frame far-infrared) luminosity of the Cloverleaf will 
still be 
dominated by the AGN. We estimate L${\rm _{Bol}}$ extrapolating from the 
observed rest frame 6$\mu$m continuum which for a mid-infrared spectrum with
weak PAH but strong continuum will be AGN dominated \citep{laurent00}. 
Using L${\rm _{Bol}\sim 10\times \nu L_\nu (6\mu m)}$ based on an 
\citet{elvis94}
radio-quiet QSO SED, the AGN luminosity is  $\sim 7\times 10^{13}$L$_\odot$. 
A similar estimate $\sim 5\times 10^{13}$L$_\odot$ is obtained from
the rest frame optical \citep[observed near-infrared;][]{barvainis95} 
continuum, tracing the AGN ionizing continuum, and the same global SED.

\citet{schweitzer06} have measured PAH emission in local QSOs and compared
the PAH to far-infrared emission ratio to that for starbursting 
ULIRGs, i.e. those among a larger ULIRG sample not showing evidence for 
dominant AGN and not having absorption dominated mid-infrared spectra.
Fig.~\ref{fig:pahtofir} places the Cloverleaf on their relation between
7.7$\mu$m PAH luminosity and far-infrared luminosity.
${\rm L(PAH)/L(FIR)}$ is 0.014 for the Cloverleaf, very close to
the mean value for the 12 starburst-dominated ULIRGs of
${\rm <L(PAH)/L(FIR)>=0.0130}$. The scatter of this relation is 0.2 in the log
for these 12 comparison ULIRGs, indicating the minimum uncertainty of 
extrapolating from the PAH to far-infrared emission. The Cloverleaf thus 
extends the relation between PAH and far-infrared luminosity for the local QSOs
and ULIRGs to $\sim$5 times larger luminosities. Its PAH emission is consistent
with an extremely luminous starburst of ULIRG-like physical conditions 
powering essentially all of the rest frame far-infrared emission. 

\citet{teplitz06} present the IRS spectrum of the lensed FIR-bright Type 2 
AGN IRAS 
F10214+4724 at similar redshift. They report a marginal feature at 
6.2$\mu$m rest wavelength which they do not interpret as PAH given the lack
of a 7.7$\mu$m maximum. Given the strength of silicate emission in this target,
PAH emission may be present in the blue wing of the silicate
feature without producing a maximum, and such a component may be suggested
by comparing their Fig.~1 with the later onset of silicate emission in 
the spectra of local QSOs. The 
tentative 6.2$\mu$m peak in IRAS F10214+4724 has similar peak height as the
Cloverleaf PAH feature, in line with our interpretation and the similar rest 
frame FIR fluxes of the two objects.

With $\sim$5-10\% of its total luminosity originating in the rest frame
far-infrared and by star formation, the Cloverleaf is within the range of 
local QSOs, and not a pronounced infrared excess object. Specifically,
its ratio of FIR to total luminosity and the ratio of rest frame 
far-infrared (60$\mu$m) to mid-infrared (6$\mu$m) continuum are about twice 
those of the \citet{elvis94} radio-quiet QSO SED. Adopting the conclusion
of \citet{schweitzer06} that star formation already dominates the FIR 
emission of local PG QSOs and considering the modest FIR `excess' of the
Cloverleaf compared to the \citet{elvis94} SED then 
suggests only a small AGN contribution to its FIR luminosity. 
Other z$\sim$2 QSOs may have lower ratios of FIR and total luminosity,
and conversely larger AGN contributions to their more modest FIR emission, 
though. After correcting for
lensing, the Cloverleaf submm flux is a factor $\sim$2 above the 
typical bright z$\sim$2 QSOs of \cite{priddey03} whose rest frame B 
magnitudes in addition are typically brighter than the delensed 
Cloverleaf.    

Submillimeter galaxies host starbursts of similar luminosity as the 
Cloverleaf, at similar redshift. \citet{lutz05} and \citet{valiante07}
have obtained IRS mid-infrared spectra of 13 SMGs with median redshift 2.8,
finding mostly starburst dominated systems. A comparison can be made between
PAH peak flux density and flux density at rest wavelength 222$\mu$m which is
obtained with minimal extrapolation from observed SCUBA 850$\mu$m fluxes.
Combining the 7.7$\mu$m feature peak of 5.1mJy (Fig.~\ref{fig:irs}) with 
a $\nu^{3.5}$ extrapolation of the SCUBA flux of \citet{barvainis02} places the
Cloverleaf at ${\rm Log(S_{PAH 7.7}/S_{222\mu m})}\sim -1.2$, near the
center of the distribution of this quantity for the SMGs of 
\citet[][their Fig.4]{valiante07}.
Like the SMGs, the Cloverleaf appears to host a scaled up ULIRG-like starburst,
but with superposition of a much more powerful AGN, also in comparison to 
the gas mass.

Tracers of high density gas, in particular HCN but also HCO$^+$ have been 
detected in a few high redshift QSOs including the Cloverleaf 
\citep{barvainis97,solomon03,riechers06}. Their ratio to 
far-infrared emission is similar to the one for Galactic dense star forming 
regions, and has been used to argue for dense, high pressure star forming 
regions dominating the far-infrared luminosity of these QSOs as well as of
local ULIRGs \citep[e.g.][]{solomon03}. Intense HCN emission is observed 
also from X-ray dominated regions close to AGN \citep[e.g.][]{tacconi94},
and there is ongoing debate as to the possible contributions of chemistry and
excitation in X-ray dominated regions,
and other effects like radiative pumping, to the emission of dense molecular
gas tracers in ULIRGs and QSOs \citep{kohno05,imanishi06,gracia06}. Unlike
HCN, PAH emission is severely reduced in X-ray dominated regions close to
AGN \citep{voit92} and provides an independent check of the effects of the AGN
on the molecular gas versus the role of the host and its star formation. 
In a scenario where XDRs dominate the strong HCN 
emission and the hosts PAH emission, reproducing the consistent 
ratios of these quantities to rest-frame far-infrared over a wide range 
of far-infrared luminosities would thus require a considerable amount of 
finetuning. In contrast, these consistent ratios are a natural implication
if all these components are dominated by ULIRG-like dense star formation.
 
Our detection of PAH emission is strong support to a scenario in which the 
Cloverleaf QSO coexists with intense star formation. Applying the
\citet{kennicutt98} conversion from infrared luminosity to star formation
rate to L$_{FIR}=5.4\times 10^{12}$L$_\odot$ suggests a star formation rate
close to 1000 M$_\odot$yr$^{-1}$, which can be maintained for a gas exhaustion 
timescale of only $3\times 10^7$yr, for the molecular gas mass inferred by
\citet{weiss03}. At this time resolution, the period of QSO activity coincides
with what likely is the most significant star forming event in the history of
the Cloverleaf host.

\acknowledgements
This work is based on observations made with the \textit{Spitzer Space Telescope}, 
which is operated by the Jet Propulsion Laboratory, California Institute of 
Technology, under a contract with NASA. Support for this work was provided by
NASA under contracts 1287653 and 1287740 (S.V.,O.S.). We thank the 
referee for helpful comments.


\clearpage

\begin{deluxetable}{lcl}
\tablecolumns{3}
\tablecaption{Cloverleaf properties}
\tablehead{
\colhead{Quantity} &
\colhead{Value}    &
\colhead{Reference}}
\startdata
Redshift z               &2.55784&\citet{weiss03}\\
Amplification $\mu_L$&11&\citet{venturini03}\\
F(PAH 6.2$\mu$m)         &$1.5\times 10^{-21}$W\,cm$^{-2}$&this work\\
F(PAH 7.7$\mu$m)         &$6.1\times 10^{-21}$W\,cm$^{-2}$&this work\\
L(PAH 7.7$\mu$m)\tablenotemark{a}&$7.6\times 10^{10}$L$_\odot$&this work\\
L(40-120$\mu$m)\tablenotemark{a} &$5.4\times 10^{12}$L$_\odot$&\citet{weiss03}\\
M(H$_2$)\tablenotemark{a}        &$3.0\times 10^{10}$M$_\odot$&\citet{weiss03}\\
L$_{Bol}$(QSO)\tablenotemark{a}  &$\sim 7\times 10^{13}$L$_\odot$&
   this work, ${\rm 10\times \nu L_\nu (6\mu m)}$\\  
\enddata
\tablenotetext{a}{Corrected for lensing amplification 11 and to our 
adopted cosmology $\Omega_m =0.3$,
$\Omega_\Lambda =0.7$ and $H_0=70$ km\,s$^{-1}$\,Mpc$^{-1}$ (D$_L$=20.96 Gpc).
}
\label{tab:quantities}
\end{deluxetable}

\clearpage

\begin{figure}
\epsscale{.80}
\plotone{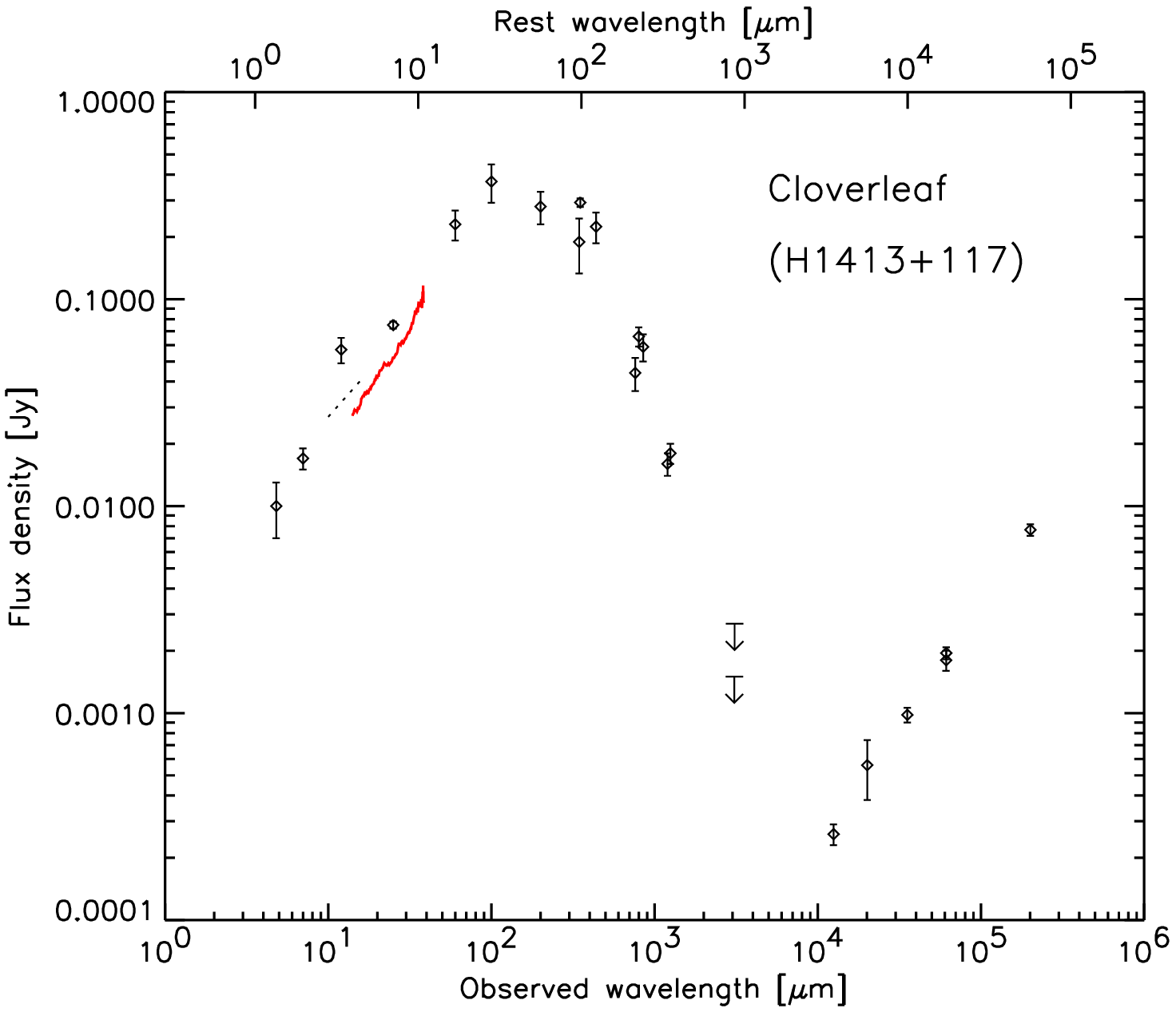}
\caption{Infrared to radio spectral energy distribution for the Cloverleaf QSO.
The IRS spectrum (continuous line) is supplemented by photometric data
from the literature \citep{barvainis95,alloin97,barvalons97,hughes97,benford99,roro00,solomon03,weiss03}. The ISOCAM-CVF spectrum of \citet{aussel98} is 
indicated by the short dotted line. The ISO 12$\mu$m flux appears too high 
while the other mid-infrared data are consistent within plausible
calibration uncertainties.}
\label{fig:sed}
\end{figure}

\clearpage

\begin{figure}
\epsscale{.80}
\plotone{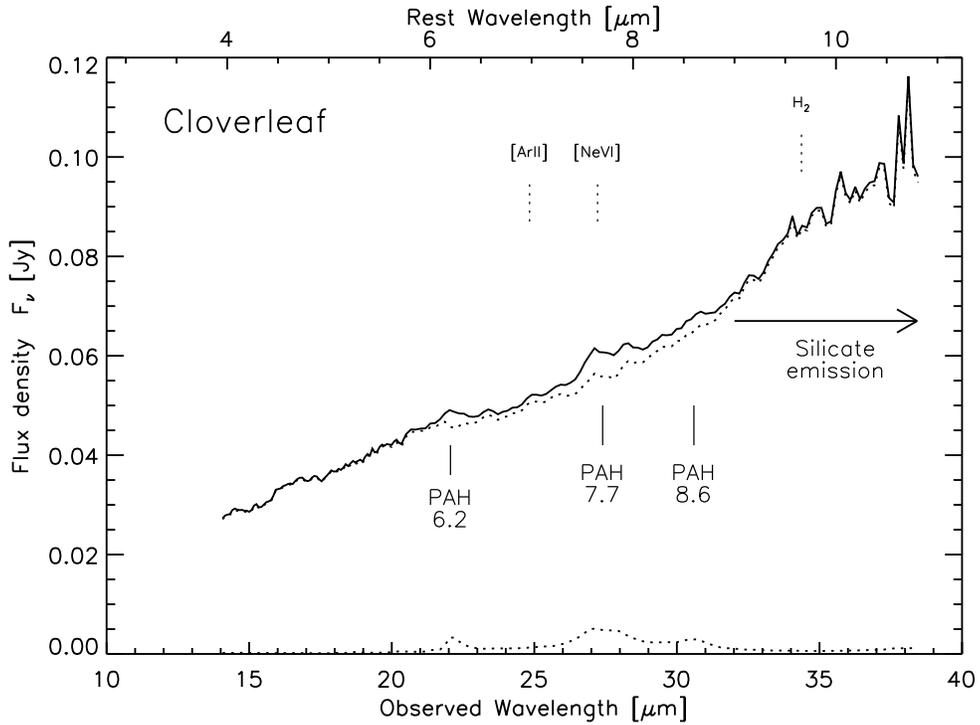}
\caption{IRS spectrum of the Cloverleaf QSO. The PAH emission features as well
as the expected locations of strong spectral lines in this wavelength range
are marked. The dotted line shows the spectrum after the subtraction of a PAH
template (spectrum of M82, see also bottom of figure), redshifted and scaled 
to the measured strength of the Cloverleaf 6.2$\mu$m PAH feature. Note that
the noise in IRS low resolution spectra increases strongly from 
$\sim$33$\mu$m towards the long wavelength end.}
\label{fig:irs}
\end{figure}

\clearpage

\begin{figure}
\epsscale{.80}
\plotone{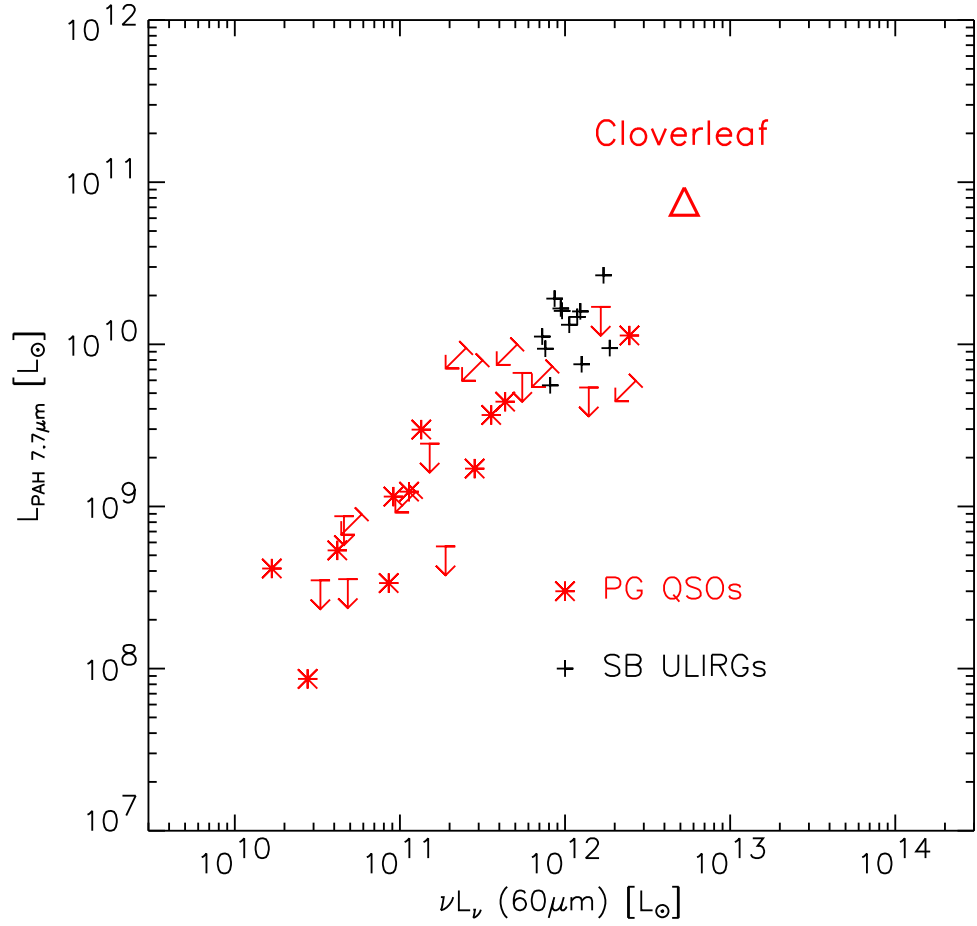}
\caption{Relation of 7.7$\mu$m PAH luminosity and rest frame FIR 
luminosity for the Cloverleaf and for local PG QSOs and 
starbursting ULIRGs from \citet{schweitzer06}.}
\label{fig:pahtofir}
\end{figure}

\end{document}